\documentclass[12pt, a4paper]{article}
\usepackage[margin=3cm]{geometry}
\usepackage[natbibapa]{apacite}

\begin{document}
\title{\vspace{-2cm}On not testing the foreign-language effect:\\
       A comment on Costa, Foucart, Arnon, Aparici, and Apesteguia~(2014)}
\author{Florian Wickelmaier\footnote{{\em Address for correspondence:}
Florian Wickelmaier, Department of Psychology, University of Tuebingen,
Schleichstrasse 4, 72076 Tuebingen, Germany. E-mail:
florian.wickelmaier@uni-tuebingen.de}}
\date{\em University of Tuebingen, Germany}

\maketitle

\begin{abstract}
In their first five studies, \citet*{CostaFoucart14} fail to provide a
statistical test of the foreign-language effect. Instead, the authors employ a
procedure in which they test the framing effects separately for the native and
the foreign language conditions. Such a procedure, however, is inappropriate
when comparing two effects; rather, a test of their difference is required.
Using the original data, it is shown that in four out of the five studies the
authors' conclusions about the existence of a foreign-language effect are
invalid.
\end{abstract}


\vspace{1cm}

In their article ```Piensa' twice: On the foreign language effect in decision
making,'' \citet*{CostaFoucart14} fail to provide a statistical test of the
foreign-language effect for their first five studies. Instead, the authors
employ a procedure in which they test the framing effects separately for the
native (NL) and the foreign (FL) language conditions. To illustrate, in their
first study the authors conclude that ``the difference between the response
distributions in the two frame versions for the FL group barely reached
significant values (Gain vs.\ Loss distribution $\chi^2(1, N = 123) = 3.7$, $p
= .05$), and was much smaller than when the task was performed in the NL (Gain
vs.\ Loss distribution $\chi^2(1, N = 124) = 14.2$, $p = .001$)''
\citep[][p.~240]{CostaFoucart14}. The procedure used by the authors, however,
does not constitute a test of their postulated hypothesis of different-sized
framing effects, and their interpretation that the former effect is smaller
than the latter is therefore not justified. This is so because the difference
between significant and not significant is not itself significant
\citep{GelmanStern06, NieuwenhuisForstmann11}. Rather, a test of the
difference of the effects is required.

To give a numerical example, consider their first study of the Asian disease
problem: When presented in the foreign language (English),
41 of
61
participants chose the sure option in the gain frame, while
31 of
62
participants chose the sure option in the loss frame. The odds ratio is
2.05.
In the native language (Spanish),
42 of
62
participants chose the sure option in the gain frame, while
21 of
62
participants chose the sure option in the loss frame. The odds ratio is
4.10.
Thus, the ratio of odds ratios across the two language conditions is
2.00.
The 95\% confidence interval for the true ratio of odds ratios is
(0.70, 5.72);
it covers one, indicating that the hypothesis of equal odds ratios cannot be
rejected with an error probability of 5\%.

\begin{table}
\caption{Odds ratios, ratio of odds ratios and 95\% confidence interval for
studies reported in \citet{CostaFoucart14}.\label{tab:OR}}
\begin{tabular}{lcccc}
\hline
      & \multicolumn{2}{c}{Odds Ratio} & Ratio of &         \\ \cline{2-3}
Study & NL & FL & Odds Ratios & 95\% CI \\ 
% latex table generated in R 3.1.2 by xtable 1.7-4 package
% Thu Jun 25 13:34:38 2015
  \hline
Asian disease (S/E) & 4.10 & 2.05 & 2.00 & (0.704, 5.724) \\ 
  Asian disease (A/H) & 4.33 & 1.59 & 2.72 & (0.602, 12.528) \\ 
  Financial crisis & 1.90 & 1.28 & 1.48 & (0.553, 3.981) \\ 
  Ticket/money lost & 2.34 & 2.29 & 1.02 & (0.377, 2.775) \\ 
  Discount & 5.61 & 1.84 & 3.05 & (0.997, 10.029) \\ 
   \hline\end{tabular}
\newline
Note: NL: native language, FL: foreign language, CI: confidence interval, S/E:
Spanish/English, A/H: Arab/Hebrew.
\end{table}

Table~\ref{tab:OR} shows the odds ratios, their ratios and confidence
intervals for the first five studies reported in
\citet{CostaFoucart14}\footnote{Code for replicating the analyses presented
here is available in the supplementary material.}. The profile likelihood
confidence intervals \citep{Agresti02} are based on the logistic regression
model
\begin{equation}
\log \frac{P(Y = 1)}{P(Y = 0)} = \beta_0 + \beta_1 X_1 + \beta_2 X_2 +
                                 \beta_3 X_1 X_2,
\label{eq:logit}
\end{equation}
where $Y = 1$ denotes that a person chose, for example, the sure option in the
Asian disease problem, $X_1 = 1$ indicates the gain version of the problem
($X_1 = 0$ the loss version), and $X_2 = 1$ the native language condition
($X_2 = 0$ the foreign language). The coefficient $\beta_3$ represents the
difference in log odds ratios, thus $\exp(\beta_3)$ is the estimated ratio of
odds ratios. Although the ratios are greater than one and point in the
hypothesized direction, the confidence intervals all include one, so none of
the effects is significant. Even when the data of the first three studies are
combined, as was done by the authors, the estimated ratio of odds ratios,
based on a model that contains the study as an additional predictor, is
1.86
(0.97, 3.56)
and not significant.

In contrast to these results, the authors' interpretation of the outcomes of
the first three studies is that ``it appears that we can safely conclude that
foreign language reduces loss aversion'' \citep[][p.~250; see their Figure~1,
p.~241]{CostaFoucart14}. They further conclude that there is a
foreign-language effect in the discount, but not in the ticket/money lost
problem (see their Table~9, p.~250). Considering the analyses presented above,
however, the conclusion rather is that there is not much evidence in favor of
a foreign-language effect in any of the five studies\footnote{In a corrigendum
to their original article, \citet{CostaFoucart15} report a meta-analysis which
shows a significant foreign-language effect based on the combined data of all
five studies. This combined effect, however, does not imply significance of
the effect for each study separately.}.

In summary, more care has to be taken when analyzing an effect that, like the
foreign-language effect, consists of the difference of two effects. Failing to
test this difference runs the risk of rendering conclusions invalid.

\bibliographystyle{apacite}
\bibliography{$HOME/w/LMU/bibtex/long,$HOME/w/LMU/bibtex/fw}

\end{document}